\documentclass[preprint,aps,showpacs]{revtex4}
\usepackage[dvips]{graphicx}

\begin{document}

\newcommand{\dfrac}[2]{\frac{\displaystyle #1}{\displaystyle #2}}
%%%%%%%%%%%%%%%%%%%%%%%%%%%%%%%%%%%%%%%%%%%%%%%%%%%%%%%%%%%%%%%%%%%%%%%%%%%%%
\preprint{VPI--IPPAP--02--02}

\title{Short Distance vs. Long Distance Physics:\\
The Classical Limit of the Minimal Length Uncertainty Relation}
\author{
S\'andor~Benczik\footnote{electronic address: benczik@vt.edu},
Lay~Nam~Chang\footnote{electronic address: laynam@vt.edu},
Djordje~Minic\footnote{electronic address: dminic@vt.edu},
Naotoshi~Okamura\footnote{electronic address: nokamura@vt.edu}, 
Saifuddin~Rayyan\footnote{electronic address: srayyan@vt.edu}, and
Tatsu~Takeuchi\footnote{electronic address: takeuchi@vt.edu}
}
\affiliation{Institute for Particle Physics and Astrophysics,
Physics Department, Virginia Tech, Blacksburg, VA 24061}

\date{April 4, 2002}

\begin{abstract}
We continue our investigation of the phenomenological implications of
the ``deformed'' commutation relations
$[\hat{x}_i,\hat{p}_j]=i\hbar[(1+\beta\hat{p}^2)\delta_{ij}
+ \beta'\hat{p}_i\hat{p}_j]$.
These commutation relations are motivated by the fact that they lead to 
the minimal length uncertainty relation which appears
in perturbative string theory.
In this paper, we consider the effects of the deformation on the 
classical orbits of particles in a central force potential.
Comparison with observation places severe constraints
on the value of the minimum length.
\end{abstract}


\maketitle
%%%%%%%%%%%%%%%%%%%%%%%%%%%%%%%%%%%%%%%%%%%%%%%%%%%%%%%%%%%%%%%%%%%%%%%%%%%%%%
\section{Introduction}

As is well known, in the case of point particles, short distance physics 
directly translates into high energy physics. 
This is a simple consequence of the Heisenberg uncertainty principle. 
In local quantum field theories, which describe the dynamics of point 
particles, the fundamental degrees of freedom are revealed at high energy,
or equivalently, at short distance. 
Also, there is a clear separation between ultraviolet and 
infrared physics from the point of view of the renormalization group.

In string theory, however, there is growing evidence that the physics 
at short distances, in contrast to local quantum field theory, 
is not clearly separated from the physics at 
long distances \cite{gross,mende,holo,adscft,ncft,dscft,dsrg}.
The fundamental formulation of this so--called UV/IR mixing,
as well as its observable consequences, are not understood at present.
Various authors have argued that some kind of UV/IR mixing is necessary 
to understand the cosmological constant problem \cite{banks,cosmoc} or the 
observable implications of short distance physics on inflationary
cosmology \cite{greene}.

Motivated by these questions, we have recently \cite{CMOT1,CMOT2} 
investigated various observable consequences of the UV/IR mixing
embodied in the ``deformed'' commutation relation \cite{Kempf:1995su}
\begin{equation}
[ \hat{x}, \hat{p} ] = i\hbar (1 + \beta \hat{p}^2) \;.
\label{Eq:Com1}
\end{equation}
This commutation relation implements the minimal length uncertainty relation
\begin{equation}
\Delta x \ge \frac{\hbar}{2}
             \left( \frac{1}{\Delta p} + \beta\,\Delta p
             \right)\;,
\label{Uncert}
\end{equation}
which appears in perturbative string theory \cite{gross,mende}.
Note the UV/IR mixing manifest in Eq.~(\ref{Uncert}): when the uncertainty
in momentum $\Delta p$ is large (UV), the uncertainty in the position 
$\Delta x$ is proportional to $\Delta p$ and is therefore also large (IR).
Note also that Eq.~(\ref{Uncert}) implies a lower bound for $\Delta x$:
\begin{equation}
\Delta x \ge \hbar\sqrt{\beta}\;.
\end{equation}
In the context of perturbative string theory, the existence of this
minimal length is tied to the fact that strings cannot probe distances 
shorter than the string length scale $\ell_S$ \footnote{%
As reviewed in Refs.~\cite{CMOT1,CMOT2}, 
the precise theoretical framework for the minimal length uncertainty 
relation is not at present understood in string theory. 
In particular, while the minimal uncertainty principle is 
based on the fact that strings cannot probe distances below $\ell_S$, 
other probes, such as $D$-branes \cite{joep},
can probe scales smaller than $\ell_S$.
In this situation, another type of uncertainty relation
involving spatial and temporal coordinates is
found to hold: $\Delta x \Delta t \sim \ell_S^2$ \cite{stu,yoneya}.
}. Thus,
\begin{equation}
\hbar\sqrt{\beta} \sim \ell_S\;. 
\end{equation}

In Ref.~\cite{CMOT1}, we determined the eigenvalues and eigenfunctions of 
the harmonic oscillator when the position and momentum obey 
Eq.~(\ref{Eq:Com1}), and studied the possible constraint 
that can be placed on $\beta$ by precision measurements on electrons 
trapped in strong magnetic fields.
Subsequently, in Ref.~\cite{CMOT2}, we pointed out that Eq.~(\ref{Eq:Com1})
implies the finiteness of the cosmological constant and a modification 
of the blackbody radiation spectrum of the cosmic microwave background.
One important observation made in Refs.~\cite{CMOT1,CMOT2}
was that various observable effects of the minimal length uncertainty 
relation are non--perturbative in the ``deformation parameter'' $\beta$
(\textit{i.e.} contain all orders in $\beta$) even though $\beta$
appears only to linear order in Eqs.~(\ref{Eq:Com1}) and (\ref{Uncert}).

In this paper, we continue our investigation and consider the effects of 
the ``deformation'' of the canonical commutation relations
on the orbits of classical particles in a central force potential.
We find that comparison with observation places a strong constraint
on the size of the minimum length. 
%In the conclusion, we also briefly contrast our
%results to past discussions on the implications of the
%minimal length uncertainty relation in string theory \cite{mende}.

\section{The Classical Limit}

In $D$-dimensions, 
Eq.~(\ref{Eq:Com1}) is extended to the tensorial form \cite{Kempf:1995su} :
\begin{equation}
[ \hat{x}_i, \hat{p}_j ]
= i\hbar( \delta_{ij}
          + \beta \hat{p}^2 \delta_{ij}
          + \beta' \hat{p}_i \hat{p}_j
        )\;.
\label{Eq:Com2}
\end{equation}
If the components of the momentum $\hat{p}_i$ are assumed
to commute with each other, 
\begin{equation}
[ \hat{p}_i, \hat{p}_j ] = 0\;,
\label{Eq:Com3}
\end{equation}
then the commutation relations among the coordinates $\hat{x}_i$ 
are almost uniquely determined by the Jacobi Identity  
(up to possible extensions) as 
\begin{equation}
[ \hat{x}_i, \hat{x}_j ]
= i\hbar\,
  \frac{(2\beta-\beta') + (2\beta+\beta')\beta\hat{p}^2}
       { (1+\beta \hat{p}^2) }
  \left( \hat{p}_i \hat{x}_j - \hat{p}_j \hat{x}_i
  \right)\;.
\label{Eq:Com4}
\end{equation}

In the classical limit,
the quantum mechanical commutator is replaced by
the Poisson bracket via
\begin{equation}
\frac{1}{i\hbar} [ \hat{A}, \hat{B} ] \Longrightarrow \{A,B\}\;.
\label{CLimit}
\end{equation}
So the classical limits of Eqs.~(\ref{Eq:Com2})--(\ref{Eq:Com4}) read
\begin{eqnarray}
\{x_i,p_j\} & = & ( 1 + \beta p^2 )\,\delta_{ij} + \beta' p_i p_j\;,\cr
\{p_i,p_j\} & = & 0 \;,\cr
\{x_i,x_j\} & = & \frac{(2\beta-\beta') + (2\beta+\beta')\beta p^2}
                       { (1+\beta p^2) }
                  \left( p_i x_j - p_j x_i \right)\;.
\label{Eq:Poisson1}
\end{eqnarray}
We are keeping the parameters $\beta$ and $\beta'$ fixed as
$\hbar\rightarrow 0$, which in the string theory context 
corresponds to keeping the string \textit{momentum} scale
fixed while the string \textit{length} scale is taken to zero.

Note that for Eq.~(\ref{CLimit}) to make sense,
the Poisson bracket must possess the same properties as the
quantum mechanical commutator, namely, it must be
anti-symmetric, bilinear, 
and satisfy the Leibniz rules and the Jacobi Identity.   
These requirements allows us to derive the general form of our 
Poisson bracket for any functions of the coordinates and momenta as
\begin{equation}
\{F,G\} = 
\left( \frac{\partial F}{\partial x_i}
       \frac{\partial G}{\partial p_j}
     - \frac{\partial F}{\partial p_i}
       \frac{\partial G}{\partial x_j}
\right) \{ x_i, p_j \}
+ \frac{\partial F}{\partial x_i}
  \frac{\partial G}{\partial x_j}
  \{ x_i, x_j \}\;,
\label{Eq:Poisson2}
\end{equation}
where repeated indices are summed.
In particular, we find that
the time evolutions of the coordinates and momenta are governed by
\begin{eqnarray}
\dot{x}_i & = & \{x_i,H\} 
\;=\; \phantom{-}\{x_i,p_j\}\,\frac{\partial H}{\partial p_j} 
    + \{x_i,x_j\}\,\frac{\partial H}{\partial x_j} \;,\cr
\dot{p}_i & = & \{p_i,H\}
\;=\; -\{x_i,p_j\}\,\frac{\partial H}{\partial x_j} \;.
\label{Eq:Poisson3}
\end{eqnarray}

This ``deformed'' version of classical mechanics is
not without its difficulties, the foremost being 
how one can construct ``canonical transformations'' which
relate the dynamical variables at one length scale to those at 
another.  For the minimal length to be a well defined length scale,
all dynamical variables at all length scales must obey
Eq.~(\ref{Eq:Poisson1}).  As a consequence, for instance,
one cannot identify the position of a composite particle with
the center of mass of its constituents.
In retrospect, it is not surprising that this difficulty
would exist given the UV/IR mixing nature of 
Eqs.~(\ref{Eq:Com2})--(\ref{Eq:Com4}) from which
Eqs.~(\ref{Eq:Poisson1}) have been derived.

We merely point out this difficulty as a caveat and 
do not attempt to propose any solution in the current paper.
Instead, we apply Eq.~(\ref{Eq:Poisson3}) 
to the motion of macroscopic objects and look for signatures of
the deformation.

\section{Motion in Central Force Potentials}

For the Hamiltonian of a particle in a central force potential,
\begin{equation}
H = \frac{p^2}{2m} + V(r)\;,\qquad r = \sqrt{x_i x_i}\;,
\end{equation}
the derivatives with respect to the coordinates and momenta are
\begin{equation}
\frac{\partial H}{\partial p_j} = \frac{p_j}{m}\;,\qquad
\frac{\partial H}{\partial x_j} 
= \frac{\partial V}{\partial r}
  \frac{x_j}{r}\;.
\end{equation}
Therefore,
the time evolutions of the coordinates and momenta in this case are
\begin{eqnarray}
\dot{x}_i & = &
  [\, 1 + (\beta+\beta')p^2 \,]\, \frac{p_i}{m}
- [\, (2\beta-\beta') + (2\beta+\beta')\beta p^2 \,]
  \left( \frac{1}{r} \frac{\partial V}{\partial r} \right)
  L_{ij}\, x_j \;,
\cr
\dot{p}_i & = &
- [\, (1+\beta p^2)\, x_i + \beta' (p\cdot x)\, p_i \,]
  \left( \frac{1}{r} \frac{\partial V}{\partial r} \right) \;,
\end{eqnarray}
where
\[
L_{ij} \equiv \frac{ x_i p_j - x_j p_i }{ (1+\beta p^2) }\;.
\]
The $L_{ij}$'s defined here are the generators of rotation:
\begin{equation}
\{ x_k, L_{ij} \} = x_i\,\delta_{kj} - x_j\,\delta_{ki}\;,\qquad
\{ p_k, L_{ij} \} = p_i\,\delta_{kj} - p_j\,\delta_{ki}\;.
\end{equation}
For motion in a central force potential, the $L_{ij}$'s are conserved due
to rotational symmetry:
\begin{equation}
\{ L_{ij}, H \} = 0\;.
\end{equation}
So is
\begin{equation}
L^2 \equiv -\frac{1}{2}L_{ij}L_{ji} 
= \frac{ p^2\,r^2 - (p\cdot x)^2 }{ (1+\beta p^2)^2 }\;.
\label{Msquared}
\end{equation}
The conservation of the $L_{ij}$'s imply that the motion of the particle
will be confined to a 2-dimensional plane spanned by the coordinate and
momentum vectors at any point in time. Therefore, without loss
of generality, we can assume that the motion is in the $x_1 x_2$-plane and
\begin{equation}
L_{12} = -L_{21} = L\;,\qquad L_{ij} = 0\quad\mbox{otherwise}\;.
\end{equation}
Then, the motion can be described by the time dependences of the distance from
the origin $r$, and the angle
\begin{equation}
\phi \equiv \tan^{-1}\frac{x_2}{x_1}\;.
\end{equation}
The equation of motion for $r$ is given by
\begin{eqnarray}
\dot{r}
& = & \frac{1}{2r}\frac{d}{dt}(r^2)
\;=\; \frac{1}{2r}\frac{d}{dt}(x_i x_i) 
\;=\; \frac{x_i}{r}\, \dot{x_i} \cr
& = & \frac{1}{m}[\,1+(\beta+\beta') p^2\,]\,p_r \;,
\label{rEq}
\end{eqnarray}
where
\begin{equation}
p_r \equiv \frac{(p\cdot x)}{r}
= \sqrt{ p^2 - \frac{L^2 (1+\beta p^2)^2}{r^2} }\;.
\end{equation}
Since the energy, $E$, is also conserved, we can write the momentum
squared as a function of $r$ via
\begin{equation}
p^2 = 2m[\,E - V(r)\,]\;.
\label{EminusV}
\end{equation}
Therefore, right hand side of Eq.~(\ref{rEq}) can be written completely
in terms of conserved quantities and functions of $r$:
\begin{equation}
\frac{dr}{dt}
= \frac{1}{m}
  \biggl[ 1+2m(\beta+\beta')( E - V )
  \biggr]
  \sqrt{ 2m( E - V ) 
       - \dfrac{ L^2 \left[ 1 + 2m\beta( E - V )
                     \right]^2 
               }
               { r^2 }
       }\;.
\label{rEq2}
\end{equation}
The equation of motion for the angle $\phi$ is
\begin{eqnarray}
\dot{\phi}
& = & \frac{x_1\dot{x_2} - x_2\dot{x_1}}{r^2} \cr
& = & \frac{L}{mr^2}
      \left\{ [\,1+(\beta+\beta')p^2 \,] (1+\beta p^2)
            + [\,(2\beta-\beta') + (2\beta+\beta')\beta p^2 \,]
              \left( mr\frac{\partial V}{\partial r} \right)
      \right\}\;.
\label{phiEq}
\end{eqnarray}
Again, using Eq.~(\ref{EminusV}), the right hand side can be
written in terms of conserved quantities and functions of $r$ only:
\begin{eqnarray}
\frac{d\phi}{dt}
& = & \frac{L}{mr^2}
  \Biggl\{ \left[\, 1 + 2m(\beta+\beta')( E - V )\,
          \right]
          \left[\, 1 + 2m\beta( E - V )\,
          \right]
  \Biggr. \cr
& & \Biggl.
        + \left[\, (2\beta-\beta')
               + 2m\beta(2\beta+\beta')( E - V )\,
          \right]
          \left(\, mr\frac{\partial V}{\partial r}\,
          \right)
  \Biggr\}\;.
\label{phiEq2}       
\end{eqnarray}
From Eqs.~(\ref{rEq2}) and (\ref{phiEq2}), we find
\begin{equation}
\frac{d\phi}{dr}
= \frac{L}{r^2}\,
  \dfrac{ 1 + 2m\beta( E - V )
        + \dfrac{ (2\beta-\beta')
                + 2m\beta(2\beta+\beta')( E - V )
                }
                { 1 + 2m(\beta+\beta')( E - V )
                }
          \left( mr\frac{\partial V}{\partial r} \right)
        }
        { \sqrt{ 2m( E - V )
        - \dfrac{ L^2 \left[ 1 + 2m\beta( E - V )
                      \right]^2 
                }
                { r^2 }
               }
        }\;.
\label{dphidr}
\end{equation}
In principle, this equation can be integrated to obtain the $\phi$ 
dependence of $r$.  We will solve Eq.~(\ref{dphidr}) for the
harmonic oscillator and Coulomb potentials in the following two
cases: 
\renewcommand{\theenumi}{\Alph{enumi}}
\begin{enumerate}
\item
$\beta\neq 0$, $\beta'=0$, in which Eq.~(\ref{dphidr}) simplifies to
\begin{equation}
\frac{d\phi}{dr}
= \frac{L}{r^2}\,
  \dfrac{ 1 + 2m\beta\left( E - V + r\frac{\partial V}{\partial r}
                     \right)
        }
        { \sqrt{ 2m( E - V ) 
        - \dfrac{ L^2 \left[ 1 + 2m\beta( E - V )
                      \right]^2 
                }
                { r^2 }
               }
        }\;,
\label{dphidr2}
\end{equation}
and 
\item
$\beta=0$, $\beta'\neq 0$, in which Eq.~(\ref{dphidr}) simplifies to
\begin{equation}
\frac{d\phi}{dr}
= \frac{L}{r^2}\,
  \dfrac{ \;\;1
        - \dfrac{ 1 }
                { (1/\beta') + 2m( E - V ) }
          \left( mr\frac{\partial V}{\partial r} \right) \;\;
        }
        { \sqrt{ 2m( E - V )
        - \dfrac{ L^2 }
                { r^2 }
               }
        }\;.
\label{dphidr3}
\end{equation}

\end{enumerate}

\section{The Harmonic Oscillator Potential}

We first consider the harmonic oscillator potential
\begin{equation}
V(r) = \frac{1}{2}m\omega^2r^2\;.
\end{equation}

\subsection{$\beta\neq 0$, $\beta'=0$ case}

For the harmonic oscillator,
Eq.~(\ref{dphidr2}) 
can be cast into the form
\begin{equation}
\frac{d\phi}{dr^2}
= \frac{1}{2}
  \left[ \frac{ r_{\max} r_{\min} }
              { r^2 \sqrt{ (r_{\max}^2-r^2)(r^2-r_{\min}^2) } }
       + \frac{ \sin\alpha        }
              {     \sqrt{ (r_{\max}^2-r^2)(r^2-r_{\min}^2) } }
  \right]\;,
\label{dphidrrHMO}
\end{equation}
where
\begin{eqnarray}
r^2_{\max/\min}
& \equiv & \frac{ E + \beta m \omega^2 L^2 ( 1 + 2m\beta E )
            \pm \sqrt{ E^2 - \omega^2 L^2 ( 1 + 2m\beta E ) }
           }
           { m\omega^2 ( 1 + \beta^2 m^2 \omega^2 L^2 ) } \cr
& = & r_{\pm}^2
      \mp \left( \frac{ 2\,r_{\pm}\,r_{\mp}^3 }{ r_+^2 - r_-^2 }
          \right) \varepsilon
      \mp \left\{ \frac{ r_{\mp}^4 ( 5 r_{\pm}^4 - 2 r_+^2 r_-^2 + r_{\mp}^4) }
                       { (r_+^2 - r_-^2)^3 }
          \right\} \varepsilon^2
    + \mathcal{O}(\varepsilon^3)\;,
\end{eqnarray}
and
\begin{eqnarray}
r_{\pm}^2 & \equiv & \left(\frac{ E\pm\sqrt{E^2 - \omega^2 L^2} }
                         { m\omega^2 }
                     \right)\;,\cr
\varepsilon & \equiv & \tan\alpha \;\equiv \; \beta m \omega L\;.
\end{eqnarray}
$r_{\max/\min}$ are the turning points when $\beta\neq 0$ and
$r_{\pm}$ are the turning points when $\beta=0$.
When $\varepsilon = \beta m\omega L$ satisfies the condition
\begin{equation}
0 
< \varepsilon
< \frac{ (r_+^2 - r_-^2)^2 }
       { 4 r_+ r_- (r_+^2 + r_-^2) }\;,
\label{betaregion}
\end{equation}
it is possible to show that
\begin{equation}
r_{-} < r_{\min} < r_{\max} < r_{+}\;.
\end{equation}
When $\varepsilon = \beta m\omega L$ 
exceeds the upper bound of the region Eq.~(\ref{betaregion}),
no solution exists.

\begin{figure}[ht]
\begin{center}
\includegraphics[scale=0.7]{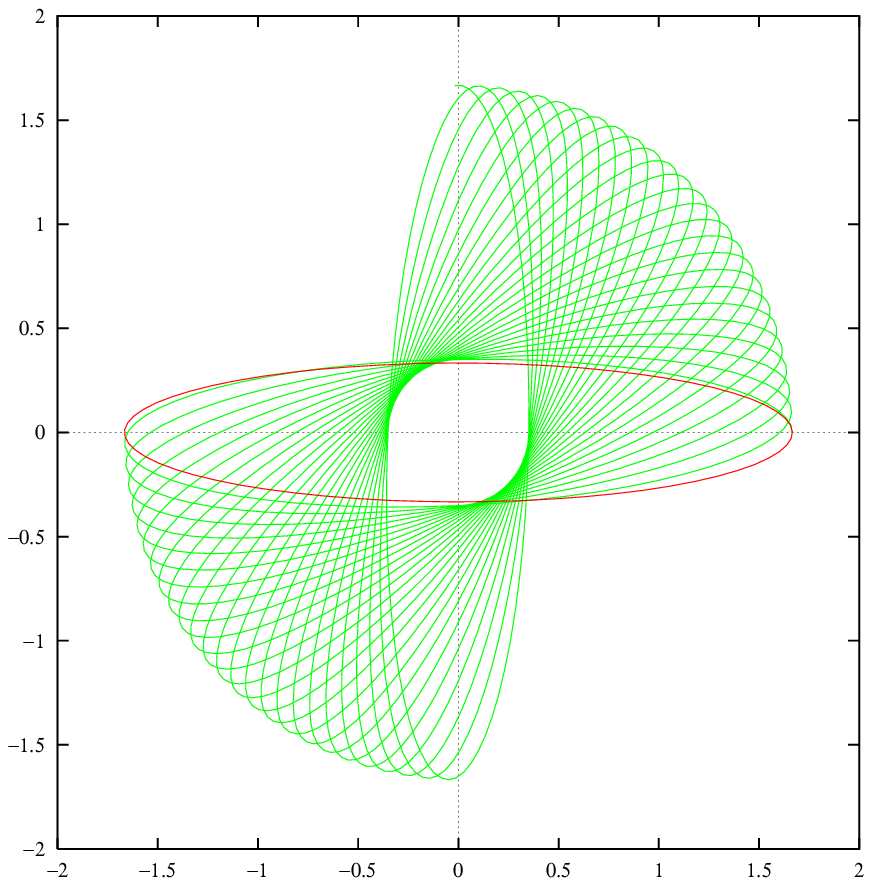}
\includegraphics[scale=0.7]{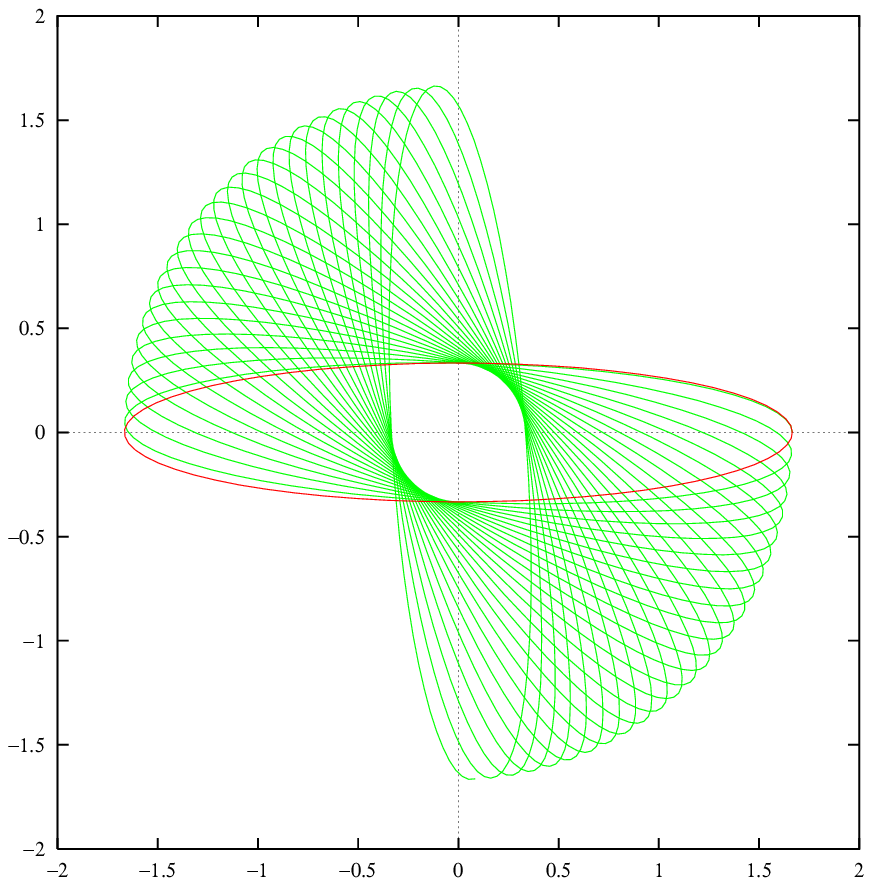}
\caption{The trajectory of a mass in a harmonic oscillator
potential with $r_+/r_- = 5$,
$\beta m\omega L = 0.01$, $\beta'=0$ (left), and
$\beta=0$, $\beta' m\omega L = 0.01$ (right).
The length scale is in units of $(r_+ + r_-)/2$.
The red line indicates the orbit when $\beta=\beta'=0$.
The motion is counter clockwise along the trajectory starting
from the aphelion on the positive $x$ axis. 
25 complete revolutions are shown.
The trajectory is precessing counter clockwise on the left,
and clockwise on the right.}
\label{Harmonic}
%\end{center}
%\end{figure}

\vspace{1cm}

%\begin{figure}[ht]
%\begin{center}
\includegraphics[scale=0.7]{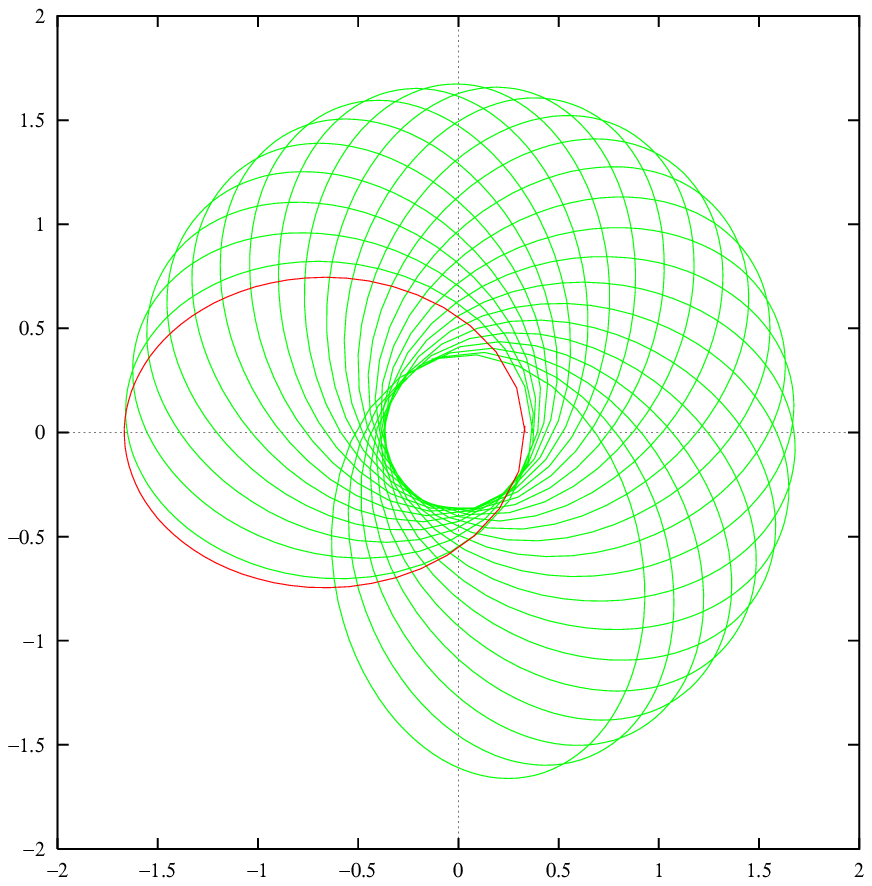}
\includegraphics[scale=0.7]{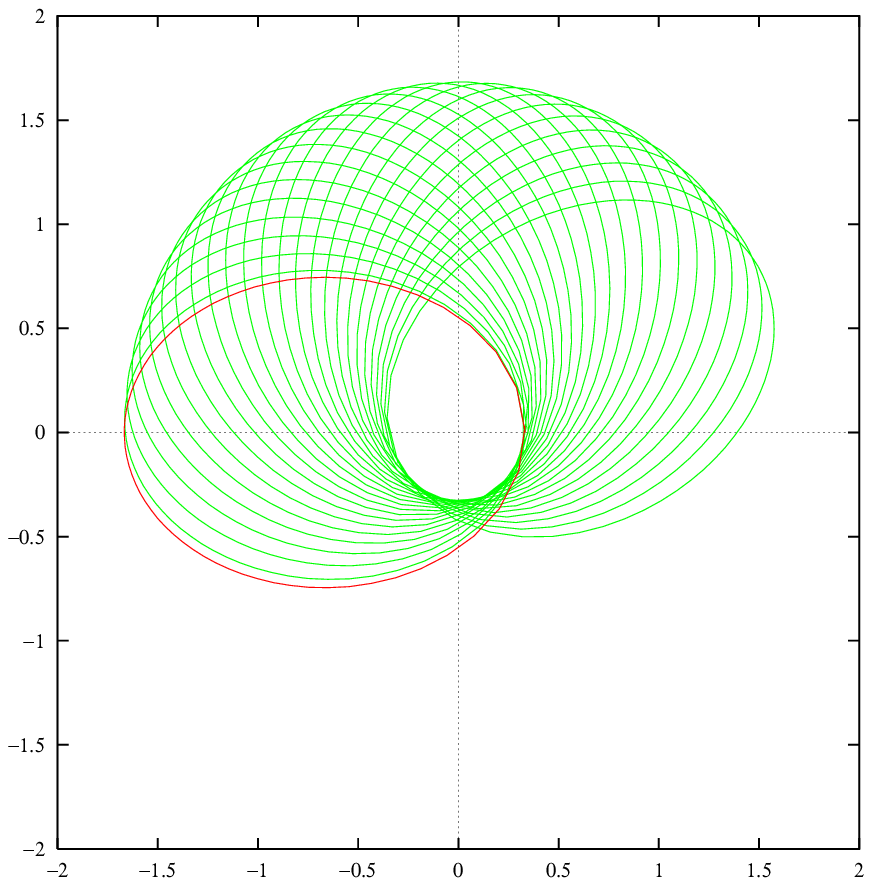}
\caption{The trajectory of a mass in a coulomb
potential with $r_+/r_- = 5$,
$2 m|E|\beta = 0.01$, $\beta'=0$ (left), and 
$\beta=0$, $2 m|E|\beta' = 0.01$ (right). 
The length scale is in units of $(r_+ + r_-)/2$.
The red line indicates the orbit when $\beta=\beta'=0$.
The motion is counter clockwise along the trajectory starting
from the perihelion on the positive $x$ axis. 
25 complete revolutions are shown. For both cases,
the trajectory is precessing clockwise.}
\label{Coulomb}
\end{center}
\end{figure}

Eq.~(\ref{dphidrrHMO}) can be integrated to yield
\begin{eqnarray}
\phi(r)
& = & \frac{1}{2}
\left[ \arcsin\left\{ \frac{ (r^2 - r^2_{\min})\, r^2_{\max}
                           - (r^2_{\max} - r^2)\, r^2_{\min}
                           }
                           { (r^2_{\max} - r^2_{\min})\, r^2 }
              \right\}
\right. \cr
& &
\left. \quad
      +\sin\alpha\,
       \arcsin\left\{ \frac{ (r^2 - r^2_{\min}) - (r^2_{\max} - r^2) }
                           { (r^2_{\max} - r^2_{\min}) }
              \right\}
\right]\;.
\end{eqnarray}
In particular, we find
\begin{equation}
\phi(r_{\max}) - \phi(r_{\min}) 
= \frac{\pi}{2}\left( 1 + \sin\alpha \right)\;,
\label{precessionHMO}
\end{equation}
which shows that the orbit will not close on itself
when $\beta\neq 0$.  It precesses by an angle of
$2\pi\sin\alpha$ per revolution. 
For $\beta\ll 1$, the precession angle is
\begin{equation}
\Delta\omega_{\beta} = 2\pi\sin\alpha \approx 2\pi(\,\beta m\omega L\,)\;.
\end{equation}
In Figure~(\ref{Harmonic}), we plot the trajectory of the
motion for a representative set of parameters.

\subsection{$\beta=0$, $\beta'\neq 0$ case}

For the harmonic oscillator,
Eq.~(\ref{dphidr3}) can be cast into the form
\begin{equation}
\frac{d\phi}{dr^2}
= \frac{ r_+ r_- }{ 2 }
  \left( \frac{1}{r^2} - \frac{1}{r_{\beta'}^2-r^2} \right)
  \frac{ 1 }{ \sqrt{ (r_+^2 - r^2)(r^2 - r_-^2) } }
\label{dphidrrHMO2}
\end{equation}
where
\begin{eqnarray}
r_{\pm}^2 & \equiv & \left(\frac{ E\pm\sqrt{E^2 - \omega^2 L^2} }
                         { m\omega^2 }
              \right)\;,\cr
r_{\beta'}^2 & \equiv & 
r_+^2 + r_-^2 + \frac{ 1 }{ m^2\omega^2\beta' }\;. 
\end{eqnarray}
Note that the turning points, $r_{\pm}$, do not depend on $\beta'$.
In the limit $\beta'\rightarrow 0$, we have 
$r_{\beta'}^2\rightarrow \infty$, and the equation for the
$\beta=\beta'=0$ case is recovered.

Eq.~(\ref{dphidrrHMO2}) can be integrated to yield
\begin{eqnarray}
\phi(r^2)
& = & \frac{1}{2}
\left[ \arcsin\left\{ \frac{ (r^2 - r_-^2)\,r_+^2 - (r_+^2 - r^2)\,r_-^2}
                           { (r_+^2 - r_-^2)\,r^2 }
              \right\}
\right. \cr
& & \quad -\sin\alpha_+ \sin\alpha_-
\left.
       \arcsin\left\{ \frac{ (r_{\beta'}^2 - r_+^2)(r^2 - r_-^2)
                           - (r_{\beta'}^2 - r_-^2)(r_+^2 - r^2) }
                           { (r_{\beta'}^2 - r^2)(r_+^2 - r_-^2) }
              \right\}
\right]\;,
\label{phirHMO2}
\end{eqnarray}
where
\begin{equation}
\tan\alpha_{\pm} \equiv r_{\pm} m\omega\sqrt{\beta'}\;.
\end{equation}
Note that $\alpha_{\pm}\rightarrow 0$ in the limit $\beta'\rightarrow 0$.
From Eq.~(\ref{phirHMO2}), we find
\begin{equation}
\phi(r_+) - \phi(r_-) 
= \frac{\pi}{2}\left( 1 - \sin\alpha_+ \sin\alpha_- \right)\;.
\label{precessionHMO2}
\end{equation}
Compared to the $\beta\neq 0$, $\beta'=0$ case, 
Eq.~(\ref{precessionHMO}),
the precession is in the opposite direction:
for each revolution, the angle swept is smaller than $2\pi$
by $2\pi \sin\alpha_+ \sin\alpha_-$.
For $\beta'\ll 1$, the precession angle is
\begin{equation}
\Delta\omega_{\beta'}
= -2\pi\,\sin\alpha_+ \sin\alpha_- 
\approx -2\pi(\,\beta' m\omega L\,)\;.
\end{equation}
In Figure~(\ref{Harmonic}), we plot the trajectory of the
motion for a representative set of parameters.

%\newpage
\section{The Coulomb Potential}

Next, we consider the attractive Coulomb potential
\begin{equation}
V(r) = -\frac{k}{r}\;,\qquad (k>0)\;.
\label{coulomb}
\end{equation}

\subsection{$\beta\neq 0$, $\beta'=0$ case}

For bound states, $E=-|E|$, Eq.~(\ref{dphidr2}) takes on the form
\begin{equation}
\frac{d\phi}{dr}
= \sqrt{r_{+}r_{-}}
  \dfrac{ 1 - \varepsilon\left\{ 1 - \frac{2(r_{+}+r_{-})}{r} \right\} }
        { \sqrt{ (r-\delta)(r-\delta^*)(r_{\max}-r)(r-r_{\min}) } }\;,
\label{dphidrCoulomb}
\end{equation}
where
\begin{eqnarray}
\varepsilon
& \equiv & 2m|E|\beta\;,\cr
r_{\pm}
& \equiv & \frac{k}{2|E|}\pm\sqrt{\frac{k^2}{4E^2}-\frac{L^2}{2m|E|}}\;,
\end{eqnarray}
and
\begin{eqnarray}
r_{\max/\min}
& = & r_{\pm}
    - \left( \frac{2 r_{\mp}^2 }{ r_{\pm} - r_{\mp} } 
      \right) \varepsilon
    - \left\{ \frac{ r_{\mp}^3 (r_{+}+r_{-})(5r_{\pm}-3r_{\mp}) }
                  { r_{\pm} (r_{\pm}-r_{\mp})^3 }
      \right\} \varepsilon^2
    + \mathcal{O}(\varepsilon^3)\;,\cr
\delta
& = & -\,\varepsilon\, (r_{+} + r_{-}) 
       \left[ 1 
            + \left\{ 1 - \frac{ 3\,(r_{+}+r_{-})^2 }
                              { 2\,r_{+}r_{-} }
              \right\} \varepsilon
            + \mathcal{O}(\varepsilon^2)
       \right] \cr
&   & + \,i\,\varepsilon^{3/2}\,\frac{(r_{+}+r_{-})^2}{\sqrt{r_{+}r_{-}}}
       \left[ 1 
            + \frac{3}{8}
              \left\{ 1 - \frac{ 7\,(r_{+}+r_{-})^2 }{ 8\,r_{+}r_{-} }
              \right\} \varepsilon
            + \mathcal{O}(\varepsilon^2) 
      \right]\;.
\end{eqnarray}
The exact forms of $\delta$ and $r_{\max/\min}$ are rather
lengthy and non-illuminating, so we will not present them here.
(See appendix~A.)
$r_{\pm}$ are the turning points when $\beta=0$, and we can see that
when $\beta>0$,
\begin{equation}
r_{-} < r_{\min} < r_{\max} < r_{+}\;,
\end{equation}
just as in the harmonic oscillator case.
The condition that $\varepsilon = 2m|E|\beta$ must satisfy for
the solution to exist is
\begin{equation}
\frac{ r_+ r_- }{ (r_+ + r_-)^2 } > 
\frac{ 8 (1 - \varepsilon)^4 }
     { 1 - 33\varepsilon - 33\varepsilon^2 + \varepsilon^3
       + (1 + 14\varepsilon + \varepsilon^2)^{3/2} }\;.
\end{equation}
Eq.~(\ref{dphidrCoulomb}) can be integrated and the solution expressed
exactly in terms of elliptic integrals. (See appendix~B.)
However, the exact expression is not particularly informative so we
present the solution to linear order in $\beta$, in which case we find
\begin{eqnarray}
\phi(r) & = & 
\left[ 1
     - \frac{(r_{\max} + r_{\min})^2}{2\,r_{\max} r_{\min}}\,\varepsilon
\right] 
\arcsin\left\{ \frac{ (r-r_{\min}) r_{\max} - (r_{\max}-r) r_{\min} }
                    { (r_{\max}-r_{\min}) r }
       \right\} \cr
& &
+ \frac{ (r_{\max} + r_{\min}) }{ r }
\sqrt{ \frac{ (r_{\max}-r)(r-r_{\min}) }{ r_{\max} r_{\min} } }
\,\varepsilon + \mathcal{O}(\varepsilon^2)\;,
\end{eqnarray}
and
\begin{equation}
\phi(r_{\max}) - \phi(r_{\min})
= \pi 
\left[ 1
     - \frac{(r_{\max} + r_{\min})^2}{2\,r_{\max} r_{\min}}\,\varepsilon
     + \mathcal{O}(\varepsilon^2)
\right] \;.
\label{precessionCoulomb}
\end{equation}
Note that, in contrast to the harmonic oscillator, the precession angle
is negative:
\begin{equation}
\Delta\omega_{\beta}
\approx -2\pi\left\{ \frac{(r_{\max} + r_{\min})^2}{2\,r_{\max} r_{\min}}\,
                     \varepsilon
             \right\}
= -2\pi\left( \frac{ 4 m|E|\beta }{ 1-e^2 } \right)\;,
\label{delta1}
\end{equation}
where $e$ is the eccentricity of the orbit.
This means that the perihelion of a planet in a gravitational
Coulomb potential will \textit{retard} instead of advance.
In Figure~(\ref{Coulomb}), we plot the trajectory of the
motion for a representative set of parameters.

%\newpage
\subsection{$\beta=0$, $\beta'\neq 0$ case}

For bound states, $E=-|E|$, Eq.~(\ref{dphidr3}) takes on the form
\begin{equation}
\frac{d\phi}{dr}
= \frac{ \sqrt{r_+ r_-} }{ 2 }
  \left( \frac{1}{r} + \frac{1}{r+r_{\beta'}}
  \right)
  \frac{ 1 }{ \sqrt{ (r_+ - r)(r - r_-) } }\;,
\label{dphidrCoulomb2}
\end{equation}
where
\begin{eqnarray}
r_{\pm}
& \equiv & \frac{k}{2|E|}\pm\sqrt{\frac{k^2}{4E^2}-\frac{L^2}{2m|E|}}\;,\cr
r_{\beta'}
& \equiv & \dfrac{ (r_+ + r_-) }
                 {\;\;\left(\dfrac{1}{2m|E|\beta'}
                      \right) - 1
             \;\;}\;.
\end{eqnarray}
As in the harmonic oscillator case, the turning points $r_{\pm}$ do not
depend on $\beta'$.
In the limit $\beta'\rightarrow 0$, we have $r_{\beta'}\rightarrow 0$,
and the equation for the $\beta=\beta'=0$ case is recovered.

Eq.~(\ref{dphidrCoulomb2}) can be integrated to yield,
\begin{eqnarray}
\phi(r)
& = & \frac{1}{2}
\left[ \arcsin\left\{ \frac{ (r - r_-)\,r_+ - (r_+ - r)\,r_- }
                           { (r_+ - r_-)\,r }
              \right\}
\right. \cr
& & \quad
\left.
+ \cos\theta_+ \cos\theta_-
       \arcsin\left\{ \frac{ (r_+ + r_{\beta'})(r - r_-)
                           - (r_- + r_{\beta'})(r_+ - r) }
                           { (r + r_{\beta'})(r_+ - r_-) }
              \right\}
\right]\;,
\label{phirCoulomb2}
\end{eqnarray}
where
\begin{equation}
\tan\theta_{\pm}
= \sqrt{ \frac{ r_{\beta'} }
              { r_{\pm}    }  
       }\;.
\end{equation}
Note that $\theta_{\pm}\rightarrow 0$ in the limit
$\beta'\rightarrow 0$.
From Eq.~(\ref{phirCoulomb2}), we find
\begin{equation}
\phi(r_+) - \phi(r_-)
= \frac{\pi}{2}\left( 1 + \cos\theta_+ \cos\theta_- \right)
= \pi\left[ 1
          - \left( \frac{1 - \cos\theta_+ \cos\theta_-}{ 2 } 
            \right)
     \right]\;.
\label{precessionCoulomb2}
\end{equation}
As in the harmonic oscillator case, the precession angle is
negative when $\beta'$ is positive.  For $\beta'\ll 1$, the precession
angle is
\begin{equation}
\Delta\omega_{\beta'}
       = -2\pi\left( \frac{1 -\cos\theta_+ \cos\theta_-}{ 2 } \right)
\approx  -2\pi\left\{ \frac{ (r_+ + r_-)^2 }{ 4\,r_+ r_- }(2m |E| \beta') 
              \right\}
= -2\pi\left( \frac{ 2m|E|\beta' }{ 1-e^2 } \right)\;.
\label{delta2}
\end{equation}
In Figure~(\ref{Coulomb}), we plot the trajectory of the
motion for a representative set of parameters.

\section{Comparison with Planetary Orbits}

Using our results, we can place constraints on $\beta$ and $\beta'$ from 
the precession of the perihelion of Mercury.
According to Ref.~\cite{Mercury}, the observed advance of the perihelion
of Mercury that is unexplained by Newtonian planetary perturbations or 
solar oblateness is
\begin{eqnarray}
\Delta\omega_\mathrm{obs}
& = & 42.980 \pm 0.002 \;\mbox{arc-seconds per century}\strut\cr
& = & \frac{ 2\pi\,(\,3.31636\pm 0.00015\,)\times 10^{-5}\,\mathrm{radians} }
           { 415.2019\;\mathrm{revolutions} } \cr
& = & 2\pi\,(\,7.98734\pm 0.00037\,)\times 10^{-8}\,
      \mathrm{radians/revolution}\;.\strut
\label{deltaObserved}
\end{eqnarray}
This advance is usually explained by General Relativity which predicts
\begin{equation}
\Delta\omega_\mathrm{GR} 
= 3\pi\left\{ \frac{ 2\,GM_\odot/c^2 }{ a\, (1-e^2) } \right\}
= 6\pi\left\{ \frac{2m|E|}{(1-e^2)\,\hbar^2} \right\}
      \left( \frac{\hbar^2}{m^2 c^2} \right)
= 6\pi\left( \frac{\lambdabar_c}{\lambdabar_d} \right)^2\;,
\end{equation}
where $2\,GM_\odot/c^2$ is the Schwarzschild radius of the Sun,
$a$ is the semi-major axis of the planet's orbit, $e$ is it's eccentricity,
and we have defined
\begin{equation}
\lambdabar_d \equiv \hbar\sqrt{ \frac{(1-e^2)}{2m|E|} }\;,\qquad
\lambdabar_c \equiv \frac{\hbar}{mc}\;.
\end{equation}
The lengths $\lambdabar_d$ and $\lambdabar_c$ are the
de Broglie and Compton wavelengths of the planet.
For Mercury, the parameters are \cite{AstroData}
\begin{eqnarray}
\frac{2\,GM_\odot}{c^2} 
    & = & 2.95325008\times 10^{3}\,\mathrm{m}\;, \cr
m   & = & 3.3022\times 10^{23}\,\mathrm{kg}\;, \strut      \cr
a \;=\; \frac{r_{\max}+r_{\min}}{2}
    & = & 5.7909175\times 10^{10}\,\mathrm{m}\;, \strut  \cr
e   & = & 0.20563069\;. \strut
%|E| & = & \frac{GM_\odot m}{2a} 
%    \;=\; 3.7839\times 10^{32}\,\mathrm{J}\;,\cr
\end{eqnarray}
Note that the product $GM_\odot$ is known to
much better accuracy than Newton's gravitational constant $G$ and
the solar mass $M_\odot$ separately.
Using these parameters we find
\begin{eqnarray}
\lambdabar_d
& = & 6.5284 \times 10^{-63}\,\mathrm{m}\;,\cr
\lambdabar_c
& = & 1.0653\times 10^{-66}\,\mathrm{m}\;,
\end{eqnarray}
and
\begin{equation}
\Delta\omega_\mathrm{GR}
= 2\pi\,(\,7.98744\times 10^{-8}\,)\,\mathrm{radians/revolution}\;.
\label{deltaGR}
\end{equation}
Comparison of Eqs.~(\ref{deltaGR}) and (\ref{deltaObserved}) yields
\begin{equation}
\Delta\omega_\mathrm{obs}-\Delta\omega_\mathrm{GR}
= 2\pi\,(\,-0.00010 \pm 0.00037\,)\times 10^{-8}\,\mathrm{radians/revolution}\;,
\label{deltaRemainder}
\end{equation}
which is consistent with zero.  
As we can see, there is not much room left for possible
extra contributions to the precession.

From Eq.~(\ref{delta1}) and (\ref{delta2}), 
the precession angle to linear order in $\beta$ and $\beta'$ is
\begin{equation}
\Delta\omega_{\beta} + \Delta\omega_{\beta'}
= -2\pi\left\{ \frac{\hbar^2(2\beta+\beta')}{\lambdabar_d^2}
       \right\}\;.
\end{equation}
The existence of a minimal length requires
\begin{equation}
\beta > 0\;,\qquad \beta + \beta' > 0\;,
\end{equation}
so we can assume that 
\begin{equation}
\Delta\omega_{\beta} + \Delta\omega_{\beta'} <
-2\pi\left( \frac{\hbar\sqrt{\beta}}{\lambdabar_d} \right)^2 < 0\;.
\end{equation}
Eq.~(\ref{deltaRemainder}) places a lower bound on 
$\Delta\omega_{\beta}+\Delta\omega_{\beta'}$ which at 
$3\sigma$ is
\begin{equation}
-2\pi\,(\,1.2\times 10^{-11}\,)\,\mathrm{radians/revolution}
< (\,\Delta\omega_{\beta} + \Delta\omega_{\beta'}\,) <
-2\pi\left( \frac{\hbar\sqrt{\beta}}{\lambdabar_d} \right)^2\;.
\label{lowerbound}
\end{equation}
Thus,
\begin{equation}
\left( \frac{\hbar\sqrt{\beta}}{\lambdabar_d} \right)^2
< 1.2\times 10^{-11}\;,
\end{equation}
or
\begin{equation}
\hbar{\sqrt{\beta}} < (3.5\times 10^{-6})\,\lambdabar_d
= 2.3\times 10^{-68}\,\mathrm{m}\;.
\label{SuperStrongConstraint}
\end{equation}
Note that this limit is 33 orders of magnitude below
the Planck length!

\section{Discussion and Conclusion}

In this paper, we have considered the effects of the minimal
length uncertainty relation on the classical orbits of 
particles in a central force potential.
Comparison with the observed precession of the perihelion of
Mercury places a strong constraint on the value of the minimum length. 

The minimal length uncertainty relation was implemented 
through the deformed commutation relation Eq.~(\ref{Eq:Com2}).
Note that even though $\beta$ and $\beta'$ appear to 
only linear order on the right hand side of Eq.~(\ref{Eq:Com2}),
our expressions for the precession angle, Eqs.~(\ref{precessionHMO}),
(\ref{precessionHMO2}), (\ref{precessionCoulomb}), and
(\ref{precessionCoulomb2}), contain all orders in $\beta$ and $\beta'$.
In that sense, our results are non-perturbative.
On the other hand, the right hand side of Eq.~(\ref{Eq:Com2}) itself
can be considered a linear approximation to a more general expression 
which leads to the minimal length uncertainty relation as discussed
by Kempf \cite{Kempf:1995su}.
This suggests that our constraint, Eq.~(\ref{SuperStrongConstraint}),
could be fairly robust.   All other possible implementation of the
minimal length uncertainty relation can be expected to lead to the
same precession of the perihelion as Eq.~(\ref{Eq:Com2}) 
to linear order in $\beta$ and $\beta'$, and result in the same
constraint on the minimal length.

The analysis of this paper based on the deformed commutation
relations can be viewed as providing a toy model 
for a full string theoretic consideration of the
implications of the minimal length uncertainty relation.
The natural question to ask is whether our constraint,
Eq.~(\ref{SuperStrongConstraint}), applies to string theory
proper or not.  This is a difficult question to answer since
the minimal length uncertainty relation is but one aspect of 
string theory, and it is not clear whether deforming the
quantum mechanical commutation relations is the correct way to
implement it.

Looking at previous works, we note that
Ref.~\cite{mende} has discussed departures from
General Relativity as implied by string theory.
These were implied both by the string theoretic modification
of Einstein's equations \cite{GSW}
\begin{equation}
R_{\mu\nu} + \frac{\alpha'}{2} 
R_{\mu\kappa\lambda\tau} R_\nu^{\kappa\lambda\tau} + \cdots = 0\;,
\label{StringEinstein}
\end{equation}
as well as the crucial distinctions between particles and strings: 
strings as extended objects do not fall freely along geodesics. 
As fundamentally extended objects (at least from the point of view 
of string perturbation theory) they are subject to tidal forces.
This leads, for example, to an energy dependent deflection angle for
the bending of light -- in clear distinction to General Relativity
in which the deflection angle is energy independent.
We have not included in our analysis any of these effects.
In particular, we have not considered possible deviations in the
background metric due to the extra terms in Eq.~(\ref{StringEinstein}).
Though the corrections to particle trajectories due to such deviations
are expected to be small, it may be worthwhile to study the problem in
more detail in light of the strong constraint we have obtained for
the minimal length.

We conclude by listing a few more caveats:
Even though our analysis is purely classical, the general 
formulation of classical systems which incorporates the classical 
limit of the minimal length uncertainty relation is not fully understood.
How one can define the ``canonical transformations'' which relate
dynamical variables at different scales while preserving the Poisson
bracket remains an open problem.
Also, the systems we considered have only a finite number of 
degrees of freedom.  It is not clear how to incorporate the effects 
of the classical limit of the minimal length uncertainty relation to 
field theory.
The classical limit of the minimal length uncertainty relation provides 
a natural generalization of the non--commutative relation between 
spatial coordinates encountered in non--commutative field theory \cite{ncft}.
What is not clear is whether the usual 
Weyl--Wigner--Moyal technology \cite{moyal} could apply even in our more 
complicated set-up, thus providing a way to analyze systems with an infinite 
number of degrees of freedom.

%%%%%%%%%%%%%%%%%%%%%%%%%%%%%%%%%%%%%%%%%%%%%%%%%%%%%%%%%%%%%%%%%%%%%%%%%%%%%%
%\newpage
\acknowledgments

We would like to thank Yasushi Nakajima, John Simonetti, 
and Joseph Slawny for helpful discussions.
This research is supported in part by a grant from the US 
Department of Energy, DE--FG05--92ER40709.

%%%%%%%%%%%%%%%%%%%%%%%%%%%%%%%%%%%%%%%%%%%%%%%%%%%%%%%%%%%%%%%%%%%%%%%%%%%%%%
%\newpage
\appendix

\section{The Turning Points for the Coulomb Potential}

The turning points for the Coulomb Potential, Eq.~(\ref{coulomb}),
are provided by the real solutions to
\begin{equation}
2m\left(-|E|+\frac{k}{r}\right)
-\frac{L^2}{r^2}
\left[ 1 + 2m\beta\left(-|E|+\frac{k}{r}\right)
\right]^2 = 0\;.
\label{A1}
\end{equation}
Defining
\begin{equation}
A \equiv \frac{k}{|E|}\;,\qquad
B \equiv \frac{L^2}{2m|E|}\;,\qquad
\varepsilon \equiv 2m|E|\beta\;,
\end{equation}
Eq.~(\ref{A1}) can be cast into the form
\begin{equation}
r^4 - A\,r^3 + B(1-\varepsilon)^2 r^2
+ 2 A B \varepsilon (1-\varepsilon)\,r
+ A^2 B \varepsilon^2 = 0\;.
\end{equation}
Since this is a quartic equation, the solutions can be obtained
algebraically (using \textit{Mathematica}) and they are:
\begin{eqnarray}
\delta\phantom{{}^*} & = & \frac{1}{4}(A-W+2iX)\;,\cr
\delta^* & = & \frac{1}{4}(A-W-2iX)\;,\cr
r_{\max} & = & \frac{1}{4}(A+W+2Y)\;, \cr
r_{\min} & = & \frac{1}{4}(A+W-2Y)\;,
\end{eqnarray}
where
\begin{eqnarray}
W & \equiv & \frac{1}{\sqrt{3}}
        \sqrt{ 3A^2 - 8 B (1-\varepsilon)^2 + 4Z
             + \frac{ 4B \left\{ B (1-\varepsilon)^4
                               + 6A^2 \varepsilon (1+\varepsilon)
                         \right\}
                    }
                    { Z }
             }  \cr
  & = & A + 4A\,\varepsilon 
          + \left( 4A - \frac{6A^3}{B} \right) \varepsilon^2 + \cdots \;, \cr
X & \equiv & \frac{1}{2}
        \sqrt{ - 3A^2 + 8 B (1-\varepsilon)^2 + W^2
             + \frac{ 2A \left\{ A^2 - 4B(1 - \varepsilon)
                                         (1 + 3\varepsilon)
                         \right\}
                    }
                    { W }
             } \cr
  & = & \frac{2A^2}{\sqrt{B}}\,\varepsilon^{3/2}
      + \frac{3A^2(8B-7A^2)}{4\sqrt{B^3}}\,\varepsilon^{5/2} + \cdots \;, \cr
Y & \equiv & \frac{1}{2}
        \sqrt{ 3A^2 - 8 B (1-\varepsilon)^2 - W^2
             + \frac{ 2A \left\{ A^2 - 4B(1 - \varepsilon)
                                         (1 + 3\varepsilon)
                         \right\}
                    }
                    { W }
             } \cr
  & = & \sqrt{A^2 - 4B} 
      - \frac{ 2(A^2-2B) }{ \sqrt{A^2 - 4B} }\,\varepsilon
      + \frac{ A^2 ( 3 A^4 - 20 A^2 B + 30 B^2) }
             { B\sqrt{(A^2 - 4B)^3} }\,\varepsilon^2 + \cdots \;, \cr
Z & \equiv & \Biggl[ \frac{1}{2}
               \Biggl\{ 2 B^3 (1-\varepsilon)^6
                     +18 A^2 B^2 \varepsilon (1+\varepsilon)
                                             (1-\varepsilon)^2
                     +27 A^4 B   \varepsilon^2
               \Biggr.
        \Biggr.  \cr
& &     \Biggl.
               \Biggl.
                     + 3\sqrt{3}A^2 B \varepsilon^{3/2}
                      \sqrt{ 27 A^4 \varepsilon
                           + 4 A^2 B (1+\varepsilon)
                                     (1-34\varepsilon+\varepsilon^2)
                           - 16 B^2 (1-\varepsilon)^4
                           }
               \Biggr\}
        \Biggr]^{1/3} \cr
  & = & B + (3 A^2 - 2B)\,\varepsilon
      + \frac{A^2\sqrt{ 3B(A^2-4B) }}{ B }\,\varepsilon^{3/2}
      + \left( 9A^2 + B - \frac{9A^4}{2B} \right)\varepsilon^2 + \cdots\;.
\label{WXYZdef}
\end{eqnarray}
In the limit $\varepsilon\rightarrow 0$, 
we recover the turning points for the $\beta=0$ case:
\begin{equation}
\delta,\delta^*\rightarrow 0\;,\qquad
r_{\max/\min}\rightarrow r_\pm = \frac{A\pm\sqrt{A^2-4B}}{2}\;.
\end{equation}

%%%%%%%%%%%%%%%%%%%%%%%%%%%%%%%%%%%%%%%%%%%%%%%%%%%%%%%%%%%%%%%%%%%%%%%%%%%%%%
%\newpage
\section{The Solution to the Coulomb Problem in terms of Elliptic Integrals}

\newcommand{\upls}{U_{\!+}}
\newcommand{\umns}{U_{\!-}}

Integration of Eq.~(\ref{dphidrCoulomb}) yields 
\begin{equation}
\phi(r) = \sqrt{r_+r_-}\,(1-\varepsilon)\,I_0 
        + 2\sqrt{r_+r_-}\,(r_++r_-)\,\varepsilon\,I_1\;,
\end{equation}
where
\begin{eqnarray}
I_0 & = & \int\frac{dr}
{\sqrt{(r-\delta)(r-\delta^*)(r_\mathrm{max}-r)(r-r_\mathrm{min})}}\;,\cr
I_1 & = & \int\frac{dr}
{r\sqrt{(r-\delta)(r-\delta^*)(r_\mathrm{max}-r)(r-r_\mathrm{min})}}\;.
\end{eqnarray}
These integrals can be expressed in terms of
the Legendre--Jacobi elliptic integrals \cite{groebner} :
\begin{eqnarray}
F(\psi, k^2) & = & \int_0^\psi\frac{d\eta}{\sqrt{1-k^2\sin^2 \eta}}\;,\cr
\Pi(\psi, \rho, k^2) & = & \int_0^\psi\frac{d\eta}
{(1+\rho\sin^2 \eta)\sqrt{1-k^2\sin^2 \eta}}\;.
\end{eqnarray}
Define
\begin{equation}
\cos\psi \equiv 
\frac{ \umns(r_\mathrm{max} - r) - \upls(r-r_\mathrm{min}) }
     { \umns(r_\mathrm{max} - r) + \upls(r-r_\mathrm{min}) }\;,
\end{equation}
with
\begin{equation}
U_\pm \equiv \sqrt{X^2 + (Y \pm W)^2}\;,
\end{equation}
and
\begin{equation}
k^2 \equiv \frac{1}{2} - \frac{W^2 + X^2 - Y^2}{2\,\upls\,\umns}\;,
\end{equation}
where $W$, $X$, and $Y$ are given in Eq.~(\ref{WXYZdef}).
The explicit expressions for the integrals are
\begin{eqnarray}
I_0(r) & = & \frac{2}{\sqrt{\upls \umns}}\ F(\psi, \ k^2) \;,\cr
I_1(r) & = &
          \frac{2}
               {\sqrt{\upls \umns}}\,
          \frac{\upls - \umns}
               {\upls r_\mathrm{min}-\umns r_\mathrm{max}}\,
          F(\psi, \ k^2) \cr
     & - &\frac{Y}
               {r_\mathrm{max} r_\mathrm{min} \sqrt{\upls \umns}}\,
          \frac{\upls r_\mathrm{min}+ \umns r_\mathrm{max}}
               {\upls r_\mathrm{min}- \umns r_\mathrm{max}}\,
          \Pi\left(\psi,
              \frac{(\upls r_\mathrm{min} - \umns r_\mathrm{max})^2}
                   {4 \upls\umns r_\mathrm{max}r_\mathrm{min}},
              k^2\right) \cr
     & + &\frac1
               {\sqrt{r_\mathrm{max} r_\mathrm{min} \delta \delta^*}}\,
          \arctan\left(\frac{Y}
                            {\sqrt{\upls \umns}}\,
                       \sqrt{\frac{\delta \delta^*}
                                  {r_\mathrm{max} r_\mathrm{min}}}\,
                       \frac{\sin \psi}
                            {\sqrt{1 - k^2 \sin^2\psi}}\right) \;.
\end{eqnarray}

%%%%%%%%%%%%%%%%%%%%%%%%%%%%%%%%%%%%%%%%%%%%%%%%%%%%%%%%%%%%%%%%%%%%%%%%%%%%%%
%\newpage

%%%%%%%%%%%%%%%%%%%%%%%%%%%%%%%%%%%%%%%%%%%%%%%%%%%%%%%%%%%%%%%%%%%%%%%%%%%%%%
\newpage

%%%%%%%%%%%%%%%%%%%%%%%%%%%%%%%%%%%%%%%%%%%%%%%%%%%%%%%%%%%%%%%%%%%%%%%%%%%%%

\end{document}